\documentclass[notoc]{JHEP3}
\usepackage{epsfig}

\title{Hybrid brane worlds in the Salam-Sezgin model}

\author{Benedict M.N. Carter\thanks{bmc55@student.canterbury.ac.nz},
Alex B. Nielsen\thanks{abn16@student.canterbury.ac.nz} and
David L. Wiltshire\thanks{david.wiltshire@canterbury.ac.nz}\\
Department of Physics and Astronomy, University of Canterbury,\\
Private Bag 4800, Christchurch, New Zealand}

\def\beq{\begin{equation}}\def\eeq{\end{equation}}
\def\bea{\begin{eqnarray}}\def\eea{\end{eqnarray}}
 \def\mrm{\mathrm} 
\def\dd{{\rm d}} \def\e{{\rm e}}\def\pt{\partial} \def\ka{\kappa}
\def\BB{{B^2}}\def\ff{f} \def\FB{{\mathbf F}} \def\DE{\Delta}
\def\ph{\phi} \def\th{\theta} \def\LA{\Lambda} \def\la{\lambda} \def\rh{\rho}
\def\al{\alpha} \def\be{\beta} \def\de{\delta} \def\ga{\gamma} 
\def\dsp{\displaystyle} \def\scr{\scriptstyle} \def\scrscr{\scriptscriptstyle}
\def\W#1{^{\raise2pt\hbox{$\scrscr#1$}}} \def\Y#1{^{\raise2pt\hbox{$\scr#1$}}}
\def\X#1{_{\lower2pt\hbox{$\scrscr#1$}}} \def\Z#1{_{\lower2pt\hbox{$\scr#1$}}}
\def\sqr#1#2#3{{\vbox{\hrule height.#2pt \hbox{\vrule width.#2pt
height#1pt\kern#1pt\vrule width.#2pt}\hrule height.#2pt}\hbox{\hskip.#3em}}}
\def\Gph{G\Z{\Phi}}  \def\qq{\nu} \def\ze{\zeta}
\def\Dal{\,{\mathchoice\sqr64{15}\sqr64{15}\sqr431\sqr331}}
\def\FF{F_{a b} F^{a b}} \def\ekph{\e^{\ka\ph}} \def\LAph{\LA\ekph}
\def\xh{x} \def\gh{\bar g} \def\lb{\bar\lambda} \def\et{\eta} \def\ch{\chi}
\def\dro{\pt_\rho} \def\Dm#1{D-#1} \def\Db#1{(\Dm#1)}
\def\gg{{\cal G}} \def\fn{\gg_n} \def\cnj{C_{n,j}}
\def\g#1{{\rm g}\Z#1} \def\xt{\widetilde{x}} \def\ee{\varepsilon}
\def\ep{\epsilon} \def\eps{\epsilon} \def\xixi#1{\left(\xi-\xi\X#1\right)}
\def\eqq{\ep\Z Q} \def\noi{\noindent} \def\CC{{\cal C}} \def\VV{{\cal V}}
\def\ekQ{\eqq Q^2\e^{2\et}} \def\elb{\lb\e^{2\ze}} \def\eLA{\LA\e^{2\ch}}
\def\eeLA{\ee\LA} \def\frac#1#2{{\textstyle{#1\over#2}}} \def\half{\frac12}
\def\BbbR{\mathbb{R}} \def\KK{Kaluza--Klein} \def\WW{{\cal W}}
\def\rB{{r\Z b}}\def\rhB{{\rh\Z b}} \def\rb{{r\X b}} \def\rhb{{\rh\X b}}
\def\rM{{r_{\lower2pt\hbox{$\!\scr-$}}}}
\def\rP{{r_{\lower2pt\hbox{$\scr+$}}}}
\def\dr{\pt_r} \def\const{\mathop{\hbox{const}}}
\def\Gkn{G_{k,n}}
\def\GL{G\Z<} \def\GR{G\Z>} \def\fu{{z}}
\def\Xnb{X_n,_b} \def\Ynb{Y_n,_b} \def\Mlow{{\cal M}\ns{low}}
 \def\rmprh{\rM^4+\rP^4\rh}
\def\rhpf{\left(1+\rh\right)}
\def\OO{{\mrm O}} \def\PP{{\cal P}}\def\QQ{{\cal Q}}
\def\fhyp#1#2#3#4{{}_2{\cal F}_1\left(\left[{#1},{#2}\right];{#3};{#4}\right)}
\def\ns#1{_{\rm#1}}
\abstract{We construct a six--dimensional warped brane world
compactification of the Salam-Sezgin supergravity model
by generalizing an earlier hybrid \KK\ / Randall--Sundrum
construction [JHEP {\bf02} (2002) 007]. In this construction the observed
universe is interpreted as a 4--brane in six dimensions, with a \KK\
spatial direction in addition to the usual three noncompact spatial
dimensions. This construction is distinct from other brane world
constructions in six dimensions, which introduce the universe as
a 3--brane corresponding to a topological defect in six dimensions, or
which require a particular configuration of matter fields on the brane.
We demonstrate that the model reproduces localized gravity on the brane in
the expected form of a Newtonian potential with Yukawa--type corrections.
We show that allowed parameter ranges include values which potentially solve
the hierarchy problem. An exact nonlinear gravitational wave solution on
the background is exhibited. The class of solutions given applies to
Ricci--flat geometries in four dimensions, and consequently includes
brane world realizations of the Schwarzschild and Kerr black holes as
particular examples. Arguments are given which suggest that the hybrid
compactification of the Salam--Sezgin model can be extended to
reductions to arbitrary Einstein space geometries in four dimensions.}

\preprint{hep-th/0602086\ \qquad JHEP {\bf07} (2006) 034}

\begin{document}
\section{Introduction}
The idea that our universe might be a surface (either a thin or thick
``brane'') embedded in a higher--dimensional spacetime with large bulk
dimensions \cite{Ak}--\cite{RS2} continues to be the focus of much interest.
While 5--dimensional models based on the Randall--Sundrum scenarios
\cite{RS1,RS2} have attracted the most attention, recently there has
been growing interest in 6--dimensional models
\cite{CN}--\cite{PST}.

One reason for investigating 6--dimensional models is to determine
whether or not some of the more interesting features of brane world
models in five dimensions are peculiar to five dimensions. Another
reason is that six dimensions allow one greater freedom in
building models with positive tension branes only~\cite{LMW}.
Possibly the strongest motivation for investigating six
dimensional models is the possibility of solving the cosmological
constant problem in a natural manner \cite{ABPQcc,Burgess}. While
codimension two branes do pose technical problems for the
cosmological constant issue \cite{CV}, which might be more easily
resolved in the model considered here, we will not address
the solution of the cosmological constant problem directly in this paper;
it remains an interesting possibility for future work.

A common feature of many of the 6--dimensional models currently
being investigated is that, in order to localize gravity on a
3--brane, a 4--brane is incorporated into the model at a finite
proper distance from the 3--brane. (See, for example, the work of
refs.\ \cite{LMW,BCCF,CDGV}, which are based on extensions of the
AdS soliton \cite{HM}.) However, due to the form of the bulk
geometry, Einstein's equations often preclude the insertion of
simple 4--branes of pure tension into these models. Several
mechanisms have been proposed to deal with this, including the
addition of a particular configuration of matter fields to the
brane \cite{CN} and ``delocalization'' of a 3--brane around the
4--brane \cite{LMW}. In this paper, we will by contrast discuss a
6--dimensional brane world model with localized gravity and a
single 4--brane with tension coupled to a scalar field,
generalizing an earlier construction by Louko and Wiltshire
\cite{LW}. The construction is fundamentally different to those
which consider our observed universe to be a codimension two defect;
in particular the physical universe is a codimension one brane in six
dimensions with an additional \KK\ direction.

The construction of ref.\ \cite{LW} was based on the bulk geometry of
fluxbranes in 6--dimensional Einstein--Maxwell theory with a bulk
cosmological constant \cite{GW}, a model which continues to attract
attention in its own right \cite{MSYK}. However, if one is interested
in 6--dimensional models then a more natural choice might be a
supersymmetric model, such as the chiral, $N=2$ gauged supergravity model of
Salam and Sezgin \cite{NS,SS}. Generally higher-dimensional models
of gravity are introduced in the context of supergravity models,
which are themselves low-energy limits of string-- or M--theory.

Supersymmetry has of course played a central role in the recently
studied codimension two brane world constructions, and the Salam--Sezgin
model has featured in the supersymmetric large extra dimensions scenario
\cite{ABPQsb,ABPQcc,ABCFPQTZ,LL,TBHA}. One motivation for providing an
alternative construction based on codimension one branes is that
discontinuities associated with codimension one surfaces in general
relativity are very well understood and easier to treat mathematically than
codimension two or higher defects \cite{GT,Ga}. While codimension two defects
can be regularised a host of technical issues are introduced when additional
matter fields are added to the brane \cite{CV,dRT}. The construction of ref.\
\cite{LW} avoids these problems. Similarly, whereas the anti--de Sitter
horizon in the bulk of the Randall--Sundrum II model \cite{RS2}
can become singular upon additional of matter fields, the
construction of \cite{LW} involves a geometry which closes in a
completely regular fashion in the bulk. Full non--linear gravitational
wave solutions were exhibited in the background of
ref.\ \cite{LW}, without additional singularities.

The biggest phenomenological problem faced by the model of \cite{LW}
was that the parameter freedom available in 6--dimensional
Einstein--Maxwell theory with a cosmological constant did not seem to allow
the proper volume of the compact dimensions to be made arbitrarily large as
compared to the proper circumference of the \KK\ circle, as would be
required for a solution of the hierarchy problem. It is our aim in this paper
to demonstrate that a supersymmetric background can solve this problem, and
that an interesting hybrid compactification without singularities arises.

The model considered in this article can therefore be viewed as a five
dimensional \KK\ universe that forms a co-dimension one surface
within a six-dimensional bulk where the codimension has a
$\mathbb{Z}_2$ symmetry across the brane which smoothly terminates in
a totally geodesic submanifold, a ``bolt'', which does not suffer a conical
defect. The topology of the solution is thus
$\mathbb{R}^{4}\times \mathrm{S}^2$. We consider the case where
there is both a magnetic flux in the bulk (the fluxbrane) and a
bulk scalar field, the potential of which is dictated by the form
of the Salam-Sezgin action. While the model can in principle support any
Einstein space, we limit most of our analysis to the case where the
4--dimensional cosmological constant is zero (i.e. the observed universe is
Minkowski) in order to solve the field equations exactly. Both the bulk
magnetic field and the bulk scalar field will impact the behaviour of
gravity on the brane and we show how one can explicitly calculate the
essential features of the gravitational potential between two test masses
on the brane. Since it is assumed that the brane will correspond to
our universe (modulo the Kaluza-Klein dimension) this will
indicate how the effects of the extra dimensions and their fields
will modify four dimensional gravity.

The paper is organized as follows. In Section 2 we introduce the
Salam-Sezgin fluxbrane solution and discuss the structure of the bulk
geometry. In Section 3 we go on to discuss junction conditions arising
from the brane and show how the position of this brane is fixed
by the bulk geometry alone. In section 4 we show how this
construction gives rise to a Newton--like gravitational law in the
brane, together with the exponential corrections expected of a model with
compact extra dimensions. While the analysis is made by analogy to the case
of a scalar propagator, the calculation contains the essential features
important to the more involved calculation for gravitational perturbations.
This is justified by the presentation of nonlinear gravitational
wave solutions in Section \ref{branewave}. The hierarchy problem is
addressed in Section \ref{hierarchy}. In Appendix \ref{sec:bh} general
arguments are presented about the extension to the case of physical
universes with the geometry of general Einstein spaces, which include the
phenomenologically interesting case of de Sitter space.

\section{Salam-Sezgin fluxbranes}
\label{sec:fluxbrane}
The bosonic sector of $N=2$ chiral Einstein-Maxwell supergravity in
six dimensions -- the Salam-Sezgin model \cite{NS,SS} -- may be
truncated to the degrees of freedom described by the action:
\beq
S=\int_{M}\dd^6 x\sqrt{-g}\left({{\cal
R}\over4\ka^2}-\frac14\pt_a\ph\,\pt^a\ph -\frac1{12}\e^{-2
\ka\ph} G_{a b c} G^{a b c} -\frac14\e^{\ka\ph}\FF -
{\LA\over2\ka^2}\e^{-\ka\ph}\right) \label{action}
\eeq
where $F_{a b}$ is the field strength of a $U(1)$ gauge field,
$G_{a b c}$ is the 3-form field strength of the Kalb-Ramond field,
${\cal B}_{ab}$, $\ph$ is the dilaton, $\ka^2=4\pi G\Z6$ and
$\LA={g\Z1}^2/(\ka^2)>0$, where $g\Z1$ is the $U(1)$ gauge
constant. Generically, the bosonic action (\ref{action}) is
supplemented by the contribution of additional scalars, $\Phi^A$,
belonging to hypermultiplets. However, these may be consistently
set to zero.

We will also make the additional simplification of setting the Kalb-Ramond
field, ${\cal B}_{ab}$, to zero, as we wish to consider just the
simplest non-trivial fluxbrane solutions in the Salam-Sezgin
model. This leaves us with the field equations:
\bea &&G_{ab}=2\ka^2\e^{\ka\ph}
\left(F_{ac}{F_b\,}^c-\frac14
g_{ab}F^{cd}F_{cd}\right)+\ka^2\left(\pt_a\ph\,\pt_b\ph-
\half\pt^c\ph\,\pt_c\ph\right)-g_{ab}\LA\e^{-\ka\ph},
\label{Fe1}
\\
&&\pt_a\left(\sqrt{-g}\e^{\ka\ph}F^{ab}\right)=0,
\label{Fe2}
\\
&&\Dal\ka\ph-\half\ka^2\e^{\ka\ph}\FF+\LA\e^{-\ka
\ph}=0.
\label{Fe3}
\eea
Static fluxbrane solutions may be found
by assuming a metric ansatz of the form
\beq
{\dd
s\Z6}^2=r^2\gh_{\mu\nu}\dd\xh^\mu\dd\xh^\nu+{\ff(r)^2\dd r^2\over\DE(r)}+
\DE(r)\dd\th^2,
\label{thickbulkG}
\eeq
where $\th$ is the Kaluza-Klein direction, $r$ is the radion and
$\gh_{\mu\nu}(\xh)$ is the metric on a $4$--dimensional Einstein spacetime
of signature $(-+++)$, such that
\beq
\bar{\cal R}_{\mu\nu}= 3\lb\,\gh_{\mu\nu}\,.
\label{thickbraneG}
\eeq
Additionally, we assume that $\ph=\ph(r)$, and that the $U(1)$ gauge
field consists purely of magnetic flux in the bulk
\bea
\FB=
{\sqrt{8} B\ff e^{-\ka\ph}\over\ka r^4}
\,
\dd r\wedge\dd\th
\,,
\label{thickF}
\eea

Rather than solving the field equations directly, fluxbrane solutions
are often conveniently obtained by double analytic continuation of
black hole solutions with a central electric charge. This double
analytic continuation technique was in fact first introduced when fluxbranes
were first constructed \cite{GW}, in $D$-dimensional Einstein-Maxwell
theory with a cosmological constant. In the present model, the
dual 6-dimensional black hole spacetime is obtained by the continuation
\beq
{\overline{\xh}}\Z0\to i\xt\Z1;
\quad
{\overline{\xh}}_i\to\xt_{i+1}\quad (i=1,2,3);
\quad
\th\to i t;
\quad
B\to -i Q;
\eeq
where it is assumed that $\pt/\pt x\Z0$ is a Killing vector and that
the Einstein space metric is written in coordinates with $\gh\Z{00}<0$
and $\gh\Z{0i}=0$.

Black hole type solutions are not well-studied in the case of the
field equations (\ref{Fe1})--(\ref{Fe3}), however, on account of the fact
that no conventional black holes exist for the Einstein-Maxwell scalar with
a Liouville potential. There are no
solutions with a regular horizon which are asymptotically flat,
asymptotically de Sitter or asymptotically anti-de Sitter \cite{PW}. There do
exist black hole type solutions with regular horizons which possess
``unusual asymptotics'' at spatial infinity \cite{PW,CHM}.
From the point of view of the double analytically dual fluxbranes, the
asymptotic structure of the black hole spacetimes is irrelevant, and
we are simply interested in the most general solution to the field
equations (\ref{Fe1})--(\ref{Fe3}) with the ansatz
(\ref{thickbulkG})--(\ref{thickF}).

To the best of our knowledge the full solutions of the field equations
(\ref{Fe1})--(\ref{Fe3}) with arbitrary $\lb$ and $\LA$ have not been
written down, either for the fluxbranes or the double analytically
dual static black hole type geometries. The $\lb=0$ case has
been given previously \cite{ABCFPQTZ}. The general case with
non-zero $\lb$ does not appear to readily yield a closed form
analytic solution. Its properties are discussed in Appendix \ref{sec:bh}.

The fluxbrane solution for $\lb=0$ takes the form
\bea
f(r)&=&r,\label{fofr}\\
\ph(r)&=&{2\over\ka}\ln(r),\label{phiofr}\\
\DE(r)&=&{A\over r^2}-{\BB\over r^6}-{\LA\over8} r^2\,,\label{trianglevar}
\eea
where we require that $A>0$ so that $\DE(r)$ has at least one root.

The finite limits of the range of the bulk coordinate $r$ are the points
at which $\DE(r)=0$, since $\DE(r)>0$ is required to preserve the
metric signature. There are at most two positive zeroes of $\DE(r)$,
located at
\beq
r_{\pm}^4={4A\over\LA}\left( 1\pm\sqrt{1-{B^2\LA\over2 A^2}}\right).
\label{rplusminus}
\eeq
For $\LA>0$, reality of $r$ implies the condition $B^2\LA\leq2 A^2$.
For $\LA<0$ there is a single positive zero of $\DE$ at $\rM$.
Our primary interest is of course for $\LA>0$, which is the case in the
Salam-Sezgin model.

We wish the geometry to be regular at points where $\DE(r)=0$ and we
therefore impose the condition that $\th$ be periodic with period
\beq
{4\pi\over\pt_r\DE(r)}\bigg|_{r\Z0}\,,
\eeq
where $r\Z0$ is any positive zero of $\DE(r)$ such that $\pt_r
\DE(r)|_{r\Z0}\neq0$.
In these circumstances the $r=r\Z0$ submanifold is totally geodesic, namely
a ``bolt'' in the terminology of \cite{GH}. In the case that two zeroes of
$\DE(r)$ exist we can fix the bolt to be at either $\rM$ or $\rP$ but not
both simultaneously as the fixing of the period of $\th$ allows
the geometry to be regular at only one of the zeroes of $\DE(r)$.

\section{Adding a thin brane}
\label{sec:thinbrane}
We now follow the construction of \cite{LW} and add a thin brane --
namely a timelike hypersurface of codimension one -- at a point $\rB$
such that $\rM\leq \rB\leq \rP$ and $\DE(\rB)>0$. To do this we
add the term
\beq
S_\mathrm{brane} = -\int
\dd^{6}x\;\de(r-\rB)\;{T\over\ka^2}\,\e^{-\la\ka\ph}
\sqrt{-h}\,,
\label{braneaction}
\eeq
to the action (\ref{action}) where $h_{ij}$ is the induced metric
on the brane, $h=\det\left(h_{ij}\right)$, Latin indices $i,j,\dots$ run
over the five dimensions on the brane $(\th,x^{\mu})$, $T$ is a nonvanishing
constant proportional to the brane tension, and $\la$ is a dimensionless
coupling constant. The tension of the thin brane is coupled to the scalar
field in order to make the scalar field equation consistent at the junction
between the two spacetimes: given that the derivative of the scalar must
be discontinuous there the boundary term which was assumed to vanish in
deriving (\ref{Fe3}) will no longer be zero.

In a Gaussian normal coordinate system for the region near the thin brane,
with normal coordinate $\dd\et=\DE^{-1/2}r\dd r$, the induced metric takes
the form
\beq
h_{ij} = g_{ij}-n_i n_j =\left(\matrix{\DE(r)&\mathbf0\cr
\mathbf0&r^2\overline{g}_{\mu\nu}\cr}
\right)\,,
\label{imetric}\eeq
where
\beq
n_i={\pt_i\eta}={r\over\sqrt{\DE}}\,\de_i^r\,.
\eeq
We then impose $\mathbb{Z}_2$ symmetry about $\rB$ by pasting a
second copy of the bulk geometry on the other side of the thin
brane. We label the bulk geometry to the left of the brane ${\cal
M}_-$, and the geometry to the right ${\cal M}_+$. A pictorial
embedding diagram of the bulk dimensions is shown in Fig.~\ref{embedfig}.

\EPSFIGURE[p]
{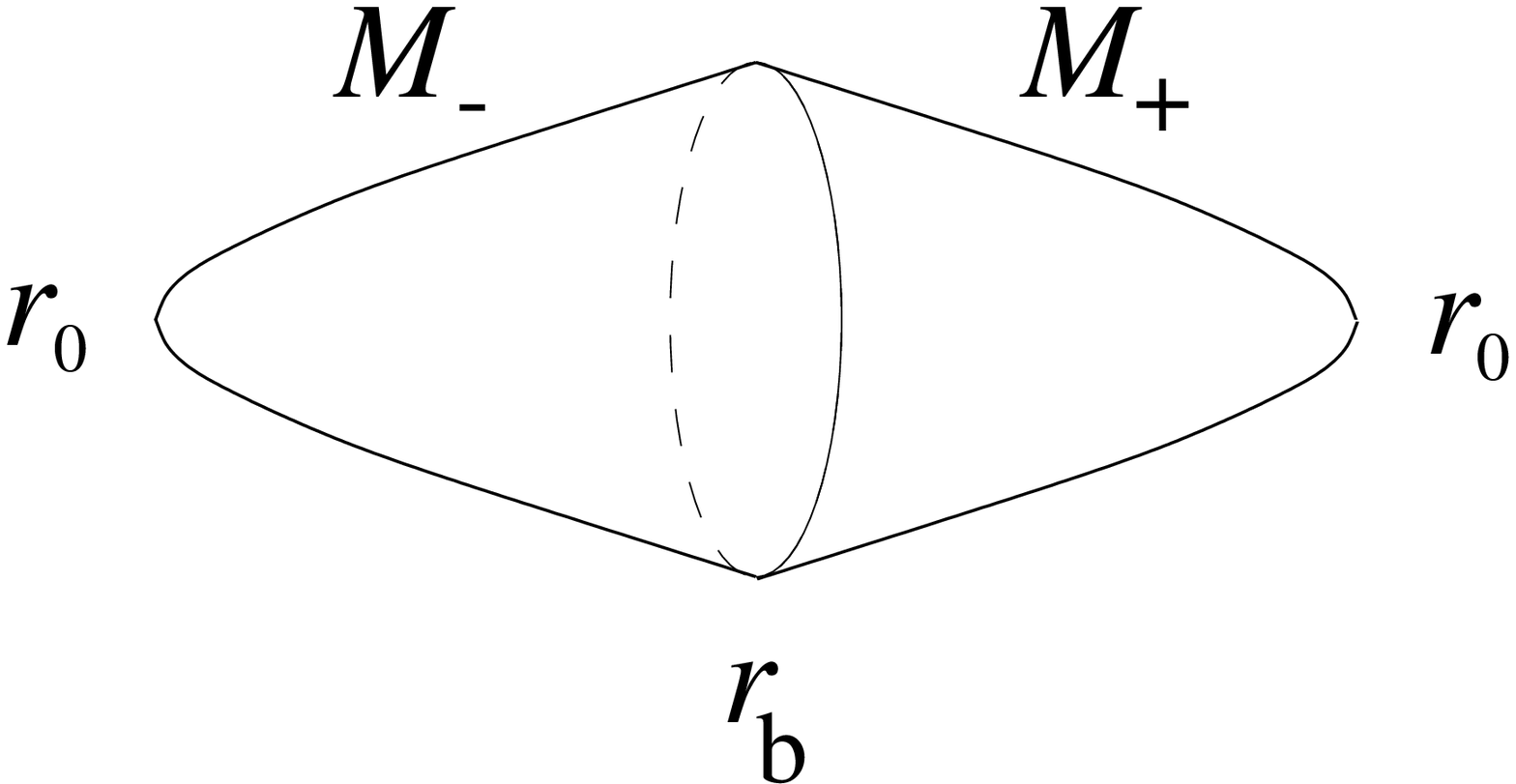,height=1.84in,width=80mm}
{An embedding of the bulk ($r,\th$) dimensions of
${\cal M}$ into~$\BbbR^3$.\label{embedfig}}

The field equations (\ref{Fe1})--(\ref{Fe3}) are modified by terms arising
from the variation of the action (\ref{braneaction}), but can still be
satisfied by appropriate junction conditions, according to the standard
thin--shell formalism. In particular, the modification to the Einstein
equation (\ref{Fe1}) is satisfied provided the discontinuity in the extrinsic
curvature,\break $K_{ij}= {h_i}^k \, {h_j}^\ell \, \nabla_k n_\ell$ is
related to the 4--brane energy--momentum, $S_{ij}$, according to
\beq
[[ K_{ij} ]] = -2\ka^2 \left( S_{ij} - {1\over4} S \; h_{ij} \right).
\label{jc1}
\eeq
Here $[[X]]$ denotes the discontinuity in $X$ across the brane. The
modified scalar equation is satisfied provided that the boundary term
arising from the discontinuity in the derivative of $\ph$ cancels the
variation of (\ref{braneaction}) w.r.t.\ $\ph$, leading to
\beq
\left[-\half\sqrt{-g}\,n^\mu\pt_\mu\ph + {\la T\over\ka}\sqrt{-h}\,
\e^{-\la\ka\ph}\right]_{\rb}=0\,.
\label{jc2}
\eeq

The $U(1)$ gauge field strength, $F_{ab}$, can be chosen to be continuous
at the junction, so that the Maxwell--type equation (\ref{Fe2}) is
automatically satisfied. As observed in \cite{LW} it should be also
possible to choose a gluing which would change the sign of $F_{ab}$
across the junction at the expense of adding a further ``cosmological
current'' action term to the brane in addition to (\ref{braneaction}).
Such a term would now involve a coupling to the scalar field, and would
therefore modify the analysis that follows. We will not pursue that option
here.

On account of (\ref{phiofr}) the solution to (\ref{jc2}) is
\beq
T=\la^{-1}\rB^{2\la-1}\,.
\label{JC3}
\eeq
This reduces the three unknown parameters, $T,\la,\rB$, to two
independent ones. Further restrictions result from (\ref{jc1}).
For a static brane in Gaussian normal coordinates,
$K_{ij}=\half{\pt h_{ij}\over\pt\eta}$. Furthermore,
while $\eta$ does not change sign across the brane, the direction of $r$
changes sign across the brane as $r$ points from $\rM$ to $\rB$, or $\rB$
to $\rP$, depending on whether the bolt is at $\rM$ or $\rP$. Thus
\beq
\eps\left(\dd r\over\dd\eta\right)^{(-)}=
-\eps\left(\dd r\over\dd\eta\right)^{(+)}={\sqrt{\DE}\over r}
\eeq
where $\eps=\textrm{sign}(r\Z b-r\Z0)$, and the superscripts $(\pm)$ refer to
the two sides of the 4--brane and are not to be confused with $r\Z{\pm}$.
Hence the jump in the extrinsic curvature is
\beq
[[ K_{ij} ]]=-\eps{\sqrt{\DE}\over r} {\pt h_{ij}\over\pt r}\bigg |_{r=\rb}
={-T\over 2}\,\e^{-\la\ka\ph}\,h_{ij}\,,
\label{ec1a}
\eeq
where we have also used $S_{ij} = -\frac{T}{\ka^2}\,\e^{-\la\ka\ph} h_{ij}$.
Using (\ref{imetric}) to substitute for the induced metric, eq.~(\ref{ec1a})
reduces to the pair of equations
\bea
\dr\DE(\rB)&=&\dsp{\eps T\over2}\,\rB\sqrt{\DE(\rB)}\,\e^{-\la\ka\ph(\rb)}
\label{JC1}\\
\sqrt{\DE(\rB)}&=&\dsp{\eps T\over4}\,\rB^2\,\e^{-\la\ka\ph(\rb)}
\label{JC2}
\eea
or equivalently
\beq
{\pt\over\pt r}\left( {\DE\over r^2}\right)\bigg{|}_{\rb}=0\,.
\label{JCde}\eeq

Solving (\ref{JCde}) we find
\beq
\rB^4={2\BB\over A}={2 \rP^4 \rM^4\over \rP^4+\rM^4}.
\label{rbsol}\eeq
We note that the brane position does not depend on the value of the
scalar potential $\LA$.
Combining (\ref{phiofr}), (\ref{JC3}) and (\ref{JC2}) we find
\beq
\la={\eps\over4}{\rB\over\sqrt{\DE(\rB)}}\,,
\eeq
and
\beq
T=4\eps\sqrt{\DE(\rB)}\,\rB^{\left(\frac{\eps\rb}{2\sqrt{\DE(\rb)}}-2\right)},
\eeq
where $\rB$ is given by (\ref{rbsol}) and $\DE(\rB)$ by
\beq
\DE(\rB)=\left(A\over2\right)^{3/2}{1\over B}\left(1-{B^2\LA\over2A}\right)\,,
\eeq
in terms of $A$, $B$ and $\LA$.
The tension is positive if we choose the bolt to be at $r\Z0=\rM$ so that
$\eps=1$. We will avoid any potential problems associated with negative
tension branes by henceforth choosing the bolt to be at $\rM$.

\subsection{Consistency conditions when adding a thin-brane}
\label{sec:consistency}
It is possible to consider models without the restrictions which we have
chosen to place on our parameters so that we could solve the bulk field
equations. If we remove the restrictions we placed on
$f,\lb,\ph$ then we may consider what would be required of
the parameters in the bulk to leave the field equations consistent upon
addition of a thin brane at a finite distance from the fluxbrane, without
explicitly solving Einstein's equations for the bulk or junction conditions.
Even solving the Einstein equations is difficult in the general case,
although we present arguments in Appendix \ref{sec:bh} that a class of
solutions does exist in phenomenologically interesting cases, such as
a positive cosmological constant, $\lb>0$, on the brane.

Consistency conditions of the form given in \cite{LMW} can be used to
further restrict the parameters of the model. Putting $D=6$, $p=3$ and
$q=4$ in equation (2.17) of \cite{LMW} and integrating over the
boundary of the internal space we get
\bea
0 & = &\oint\dd r\,\dd\th\;r^{\al+2}\Bigg(\al\bar{\cal R}r^{-2}+(3-\al)
\tilde{\cal R}-(\al + 3)\LA\ns{Bulk}\nonumber\\ & &-\ka^{2}\bigg[(9-\al)
{T\over\ka^2}\,\e^{-\la\ka\ph}\de(r-\rB)-(3-\al){\cal T}^{\mu}_{\mu}-3
(\al-1){\cal T}^{m}_{m}\bigg]\Bigg).
\label{branesumrule}
\eea
Equations (\ref{thickbulkG}) and (\ref{thickbraneG}) give
\bea
\bar{\cal R} & = & 12\lb\,,\label{rbar}\\
\tilde{\cal R} & = & {\dr f\dr\DE-f\dr\dr\DE\over f^{3}}\,,
\label{rtilde}
\eea
and from (\ref{thickF})
\bea
{\cal T}^{\mu}_{\mu} & = &
{8B^{2}\over\ka^{2}r^{10}f^{2}}\ekph -{(\dr\ph)^{2}\DE\over f^{2}}
-{2\over\ka^{2}}\LAph\,,\\
{\cal T}^{m}_{m} & = &\ekph {4B^{2}\over\ka^{2}r^{10}f^{2}}
-\ekph {\LA\over\ka^{2}}\,.
\label{branesumse}
\eea
Putting $\al=3$ and $\LA\ns{Bulk}=0$ in (\ref{branesumrule}) this becomes
\beq
0=\oint\dd r\,\dd\th\;r^{5}\left({36\lb\over r^{2}}-6T\,\e^{-\la\ka\ph}
\de(r-\rB)+{24B^{2}\ekph\over r^{10}f^{2}}-6\LAph\right)
\label{branesumalp3}
\eeq
By examining the signs of the various terms we can find the parameter
restrictions given in Table \ref{taboo}.
\TABLE{
\begin{tabular}[h]
{c|c@{\hskip1cm}c@{\hskip1cm}c}
& $\LA<0$ & $\LA=0$ & $\LA>0$\\
\hline $\vphantom{\vbox to14pt{\vfil}}
\lb<0$ & none & $B\ne0$ or $T<0$ & $B\ne0$ or $T<0$\\
$\lb=0$ & $T>0$ & $T>0$ and $B\ne0$ & $B\ne0$ or $T<0$\\
$\lb>0$ & $T>0$ & $T>0$ & none\\
\end{tabular}
\caption{Restrictions on $B$ and the sign of $T$ for given $\lb$ and~$\LA$.}
\label{taboo}}
Some further small restrictions on the value of $B$ may result
from the junction conditions once the function $\DE(r)$ is specified for a
particular geometry, as occurs in the analogous case of ref.\ \cite{LW}.
However, as we are only able to specify an exact $\DE(r)$ in the $\lb=0$ case
(\ref{trianglevar}), and not in the general case $\lb\ne0$ case, we have not
considered these additional restrictions in Table \ref{taboo}.

\subsection{Further extensions of hybrid compactifications}
\label{sec:morehybrid}

One important question which is beyond the scope of our present analysis
is the issue of how the inclusion of extra metric degrees of freedom,
$\left\{{}^6\!g_{\mu\th},{}^6\!g_{r\th}\right\}$, in addition to the metric
components (\ref{thickbulkG}) in six dimensions, would manifest themselves
in the 4--dimensional effective theory. In the 5--dimensional
Randall--Sundrum scenario \cite{RS1,RS2} standard
model gauge fields are included by adding sources which are confined to
the brane, as distributional terms in the 5--dimensional action.

In the present model, the situation is different since the $\th$ coordinate
is understood to parameterise a regular Kaluza--Klein type direction in the
6--dimensional theory. Thus it would appear we have the freedom to multiply the
existing terms in (\ref{thickbulkG}) by appropriate functions of a new
5--dimensional scalar field, $\psi(x^\mu,r)$, and to interpret additional
non--zero components $\left\{{}^6\!g_{\mu\th}(x^\mu,r),{}^6\!g_{r\th}(x^\mu,r)
\right\}$ of the metric in terms of functions of $\psi$ multiplied by a
5--dimensional $U(1)$ gauge potential.
In this way, one would obtain a dimensional reduction from six to five
dimensions similar to a standard Kaluza--Klein reduction of the
Einstein--Hilbert action, but which is greatly complicated by the additional
non-zero matter fields of the action (\ref{action}) which contribute to
the fluxbrane background. Addition of the thin brane, to achieve the
final reduction to four dimensions, would then necessitate the addition of
further terms to (\ref{braneaction}) to satisfy the junction conditions.

The extension of the model in this fashion is of considerable phenomenological
interest, since we have the possibility not only of obtaining a massless
graviton, but also a massless photon as an analogue of the Randall--Sundrum
mode, together with a tower of very massive states.
Given the considerable amount of work involved, we will not study the question
of the existence of a massless $U(1)$ mode in four dimensions in the
present paper, nor will we calculate mass gaps for the scalar, vector and
tensor sectors. These questions are left to future work.
Our primary interest in the present paper is to check that an appropriate
graviton zero mode exists, with phenomenologically realistic corrections to
Newton's law, for parameter values which at the same time allow a solution
of the hierarchy problem. To this end we will now solve an analogous problem
for the static potential of a massless scalar field, and exhibit a
non-linear gravitational wave solution with spectral
properties equivalent to the massless scalar.

\section{Static potential of the massless scalar field}
\label{sec:Potential}
The phenomenologically important derivation of the Newtonian limit and
corrections should ideally be conducted in the context of a full
tensorial perturbation analysis about the background solution. However,
as was observed by Giddings, Katz and Randall \cite{GKR} in the case of
the Randall Sundrum model, if one is just interested in the static
potential the relevant scalar gravitational mode shares the essential
features with the static potential of a massless scalar field on the
background. This approach was adopted in \cite{LW}. In this
section we will perform a similar analysis for the background geometry
described by eqs.\ (\ref{thickbulkG})-(\ref{trianglevar}) in the
case that $\gh_{\mu\nu}=\eta_{\mu\nu}$.

\subsection{Scalar propagator\label{scalprop}}
Our calculation will closely follow that of section $5$ of ref.\ \cite{LW}.
We will add a massless minimally coupled scalar field, $\Phi$, to the
model, with action
\beq
S_\Phi=-\half\int_{{\cal M}_-}\dd^{6}x\,\sqrt{-g}\, (\nabla_a\Phi)
(\nabla^a\Phi)\,.
\eeq
on ${\cal M}_-$. This additional field $\Phi$ should not be confused with the
scalar field, $\ph$, of the Salam-Sezgin model (\ref{action}).
We will calculate the static potential of a scalar field, $\Phi$,
between two points on the thin brane with fixed $\th$.

Rather than continuing with the coordinate basis of (\ref{thickbulkG})
it is convenient to introduce a new radial coordinate $\rh$ by
\beq
\rh={r^4 -\rM^4\over \rP^4-r^4}\,,
\label{rhodef}\eeq
which maps the interval $\rM<r<\rB$ to the interval $0<\rh<\rhB$, where
\beq
\rhB=\left({\rM\over \rP}\right)^4\,,
\label{rho_b}\eeq
is the position of the brane. In terms of the new radial coordinate $\rh$
the metric (\ref{thickbulkG}) becomes,
\beq
{\dd s\Z6}^2=r^2\gh_{\mu\nu}\dd\xh^\mu\dd\xh^\nu+{r^2\rh^{-1}\,\dd\rh^2
+\ga^2r^{-6}\rh\,\dd\th^2\over2\LA\rhpf^2}\,,
\eeq
where by inverting (\ref{rhodef})
\beq
r^2=\sqrt{\rmprh\over1+\rh}\,,
\eeq
and the constant $\ga$ is defined by
\beq
\ga={\LA\over2}(\rP^4-\rM^4)=4A\sqrt{1-{\LA B^2\over2A^2}}\,.
\eeq
We shall only be interested in the case of a flat lower--dimensional
metric, $\gh_{\mu\nu}=\eta_{\mu\nu}$, in what follows.

The scalar Green's function, $\Gph$, is determined by the solution of
the massless Klein-Gordon equation,
\beq
\nabla_a\nabla^a\Gph= {1\over\sqrt{-g}}\pt_{a}
(\sqrt{-g}g^{ab}\pt_{b}\Gph) ={\de(\rh-\rh')
\de(\th-\th')\de^4(x-x')\over\sqrt{-g}}\,.
\label{KleinGordon}
\eeq
To simplify this problem we make the Fourier decomposition
\beq
\Gph(x,\rh,\ph;x',\rh',\ph')
=\int{\dd^4 k\over{(2\pi)}^5}\e^{ik_\mu(x^\mu-{x'}\vphantom{\scr x}^\mu)}\;
\sum_{n=-\infty}^{\infty}\e^{in(\th-\th')}\Gkn(\rh,\rh')
\label{Gphi}\eeq
where the indices of $k_\mu$ are raised and lowered by $\eta_{\mu\nu}$.
We substitute (\ref{Gphi}) in (\ref{KleinGordon}) to obtain
\beq
\left\{\dro\left(\rh\dro\right)+{q^2\over2\LA\rhpf^2}
-{n^2\over\rh}\left(\be+{2\rh\over\LA(1+\rh)}\right)^2\right\}
\Gkn(\rh,\rh')={\de(\rh-\rh')
\over\ga}\,.\label{sl}
\eeq
where $q^2=-k_\mu k^\mu=k_0^2-\mathbf{k}^2$, and the constant $\be$ is
defined by
\beq
\be={\rM^4\over\ga}={2\rhB\over\LA(1-\rhB)}\,.
\label{bega}\eeq
When $\rh\ne\rh'$ (\ref{sl})
is a Sturm-Liouville equation, with the general solution
\beq
\Gkn=C_n X_n(\rh)+D_n Y_n(\rh),
\eeq
where $C_n$ and $D_n$ are constants and for $n=0$,
\beq
X\Z0=\sqrt{2\over\ga}\,\PP_{\qq-1}\left(1-\rh\over1+\rh\right),\qquad
Y\Z0=\sqrt{2\over\ga}\,\QQ_{\qq-1}\left(1-\rh\over1+\rh\right),
\label{exact0}\eeq
$\PP$ and $\QQ$ being Legendre functions of the first and
second kind respectively, while for $n\ne0$,
\bea
X_n&=&{\rh^{n\be}\rhpf^{1-\qq}\over\sqrt{2n\be\ga}}\;
\fhyp{1-\qq+{2n\over\LA}+2n\be}{1-\qq-{2n\over\LA}}{1+2n\be}{-\rh}\,,
\nonumber\\ &=&{\rh^{n\be}\rhpf^{2n/\LA}\over\sqrt{2n\be\ga}}\;
\fhyp{\qq-{2n\over\LA}}{1-\qq-{2n\over\LA}}{1+2n\be}{\rh\over1+\rh}\,,
\label{Xhyp}\eea
and
\bea
Y_n&=&{\rhpf^{1-\qq}\over\sqrt{2n\be\ga}\,\rh^{n\be}}\;
\fhyp{1-\qq+{2n\over\LA}}{1-\qq-{2n\over\LA}-2n\be}{1-2n\be}{-\rh}\,,
\nonumber\\ &=&{\rhpf^{2n(\be+1/\LA)}\over\sqrt{2n\be\ga}\,\rh^{n\be}}\;
\fhyp{\qq-{2n\over\LA}-2n\be}{1-\qq-{2n\over\LA}-2n\be}
{1-2n\be}{\rh\over1+\rh}\,,\nonumber\\
\label{Yhyp}\eea
where $_2{\cal F}_1$ is a standard hypergeometric function \cite{AbS},
and for all values of $n$, including $n=0$,
\beq
\qq\equiv\half\left(1+\sqrt{1+{2q^2\over\LA}+{16n^2\over\LA^2}}\,\right)\,.
\label{nuq}
\eeq

For $n\neq0$, as $\rh\to0$, the leading two terms in the series expansions
for $X_n(\rh)$ and $Y_n(\rh)$ match those of Bessel functions,
$J_{\pm2n\be}$, or modified Bessel functions, $I_{\pm2n\be}$ in the
argument $\left|{2\over\LA}(8n^2\be-q^2)\rh\right|^{1/2}$ up to an overall
constant of proportionality:
\bea
X_n(\rh)&=&{\rh^{n\be}\over\sqrt{2n\be}}
\left[1+{8n^2\be-q^2\over2\LA(1+2n\be)}\rh+\OO(\rh^2)\right]
\nonumber
\\ &\propto&{\cases{J_{2n\be}\left(\sqrt{{2\over\LA}(q^2-8n^2\be)
\rh}\,\right)+\OO(\rh^{2+n\be}),&$n^2<q^2/(8\be)$\cr
I_{2n\be}\left(\sqrt{{2\over\LA}(8n^2\be-q^2)\rh}\,
\right)+\OO(\rh^{2+n\be}),&$n^2>q^2/(8\be)$\cr}}\ ,
\label{limit1}\eea
and
\bea
Y_n(\rh)&=&{\rh^{-n\be}\over\sqrt{2n\be}}
\left[1+{8n^2\be-q^2\over2\LA(1-2n\be)}\rh+\OO(\rh^2)\right]
\nonumber
\\ &\propto&{\cases{J_{-2n\be}\left(\sqrt{{2\over\LA}(q^2-8n^2\be)
\rh}\,\right)+\OO(\rh^{2-n\be}),&$n^2<q^2/(8\be)$\cr
I_{-2n\be}\left(\sqrt{{2\over\LA}(8n^2\be-q^2)\rh}\,
\right)+\OO(\rh^{2-n\be}),&$n^2>q^2/(8\be)$\cr}}\ .
\label{limit2}\eea
Using (\ref{exact0}), (\ref{limit1}) and (\ref{limit2}) the Wronskian of
the linearly independent solutions satisfies
\beq
\WW\left[X_n(\rh),Y_n(\rh)\right]\equiv X_n\pt_\rh Y_n-Y_n\pt_\rh X_n=
{-1\over\ga\rh}\,.
\label{Wron}
\eeq
The overall coefficients in (\ref{exact0}), (\ref{Xhyp}) and (\ref{Yhyp})
were chosen to make the r.h.s.\ of (\ref{Wron}) independent of $n$.

\subsection{Boundary and matching conditions}
We now wish to solve the inhomogeneous version of (\ref{sl}).
Without loss of generality, we pick the brane to be to a distance
$\xi=\rhB-\rh'>0$ to the right of the discontinuity.
We will later let $\xi\to0$ so that the brane explicitly becomes the
source of the discontinuity. We have general solutions to the left and right
of the discontinuity at $\rh=\rh'$, labelled
\beq
\Gkn=\cases{\GL(\rh,\rh')=A\Z1(\rh')X_n(\rh)+A\Z2(\rh')Y_n(\rh),&$\rh<\rh'$,
\cr \GR(\rh,\rh')=B\Z1(\rh')X_n(\rh)+B\Z2(\rh')Y_n(\rh),&$\rh>\rh'$,\cr}
\eeq

We will assume that $\Gkn(\rh,\rh')$ is finite at $\rh=0$,
and adopt a Neumann boundary condition at the brane $\rh=\rhB$,
\beq
\dro\Gkn|\Z{\rh=\rhb} = 0\,.
\label{neumannBC}
\eeq
Imposition of regularity of the solution as $\rh\to 0$ excludes $Y_n$
as a solution, leading to the choice $A_2(\rh')=0$.
Furthermore, (\ref{neumannBC}) applied to $\GR(\rh,\rh')$ implies
\beq
{B_2(\rh')\over B_1(\rh')}=-{\Xnb\over\Ynb}\,.
\eeq
where
\beq
\Xnb\equiv\left.\pt_\rh X_n\right|\Z{\rh=\rhb},\qquad
\Xnb\equiv\left.\pt_\rh Y_n\right|\Z{\rh=\rhb}.
\eeq

The matching conditions at $\rh=\rh'$ are
\bea
\left.\left(\GL-\GR\right)\right|\Z{\rh=\rh'}&=&0\,,\\
\left.\dro\left(\GR- \GL\right)\right|\Z{\rh=\rh'}&=&{1\over\ga\rh'}\,.
\eea

Combining the boundary conditions, the matching conditions and (\ref{Wron})
we find the solution of the boundary value problem,
\beq
\Gkn(\rh,\rh')=\cases{
\GL(\rh,\rh')=\dsp{X_n(\rh)\over\Xnb}\left(X_n(\rh')\Ynb-Y_n(\rh')\Xnb\right),&
$\rh<\rh'$,\cr
\GR(\rh,\rh')=\dsp{X_n(\rh')\over\Xnb}\left(X_n(\rh)\Ynb-Y_n(\rh)\Xnb\right),&
$\rh>\rh'$.\cr}
\label{Gfn}\eeq
The scalar Green's function, $\Gph$, is now determined by substituting
(\ref{Gfn}) into (\ref{Gphi}) and prescribing the integration
as desired at the poles, which in this case correspond to the zeroes of
$\Xnb$.

\subsection{Static potential on the brane}
Given the brane is the source of the discontinuity for the Green's
function, we set $\rh'=\rh_{b}$ by letting $\xi\to0$.
With $\rh=\rhB$, and using (\ref{Wron}), the Green's function (\ref{Gfn})
reduces to
\beq
\Gkn(\rhB)
=-{X_n(\rhB)\over\ga\rhB\Xnb}\,.
\label{braneG}\eeq

To obtain the static potential, we explicitly integrate the retarded
Green's function (\ref{Gphi}), (\ref{braneG}) over the time
difference, $t-t'$.
We note that $\Gph$ is non-zero only for $t-t'>0$, so multiplying
it by $\th(t-t')$ leaves it unchanged. We can then perform the
integration over $t$ to find
\beq
V_\Phi(\mathbf{x},\phi;\mathbf{x}',\phi') =-\sum_{n=-\infty}^{\infty}
\e^{in(\th-\th')}\int{\dd^{3}\mathbf{k}\over(2\pi)^{5}}
\,\e^{i\mathbf{k}\cdot(\mathbf{x}-\mathbf{x}')}
\int_{-\infty}^{\infty}{\dd k^0\over i(k^0-i\eps)}
{X_n(\rhB)\over\ga\rhB\Xnb},
\label{Gret}
\eeq
We are interested in the retarded Green's function, which requires that we
perform the $k^0$ integral by a contour integration with $q^2\to(k^0+i\eps)^2
-{\bf k}^2$, $\eps\to0_+$. We close the contour in the upper half plane,
to avoid the poles which correspond to the zeroes of $\Xnb$ on the real line,
and which are moved below the real line by the $\eps$--procedure.
The only residue is then due to the simple pole at $k^0=0$, and the
integration yields
\bea
V_\Phi(\mathbf{x},\phi;\mathbf{x}',\phi') &=&-\sum_{n=-\infty}^{\infty}
\e^{in(\th-\th')}\int{\dd^{3}\mathbf{k}\over(2\pi)^{4}}
\,\e^{i\mathbf{k}\cdot(\mathbf{x}-\mathbf{x}')}\,\fn(-k^2),\nonumber\\
&=&-\sum_{n=-\infty}^\infty
\e^{in(\th-\th')}\int_0^\infty{\dd k\over4\pi^3}
\,{k\sin(k|\mathbf{x}-\mathbf{x}'|)\over |\mathbf{x}-\mathbf{x}'|}
\,\fn(-k^2)\,,\nonumber\\
&=&-\sum_{n=-\infty}^\infty
\e^{in(\th-\th')}\hbox{Im}\;\left(\int_{-\infty}^\infty{\dd k\over8\pi^3}
{k\,\e^{ik|\mathbf{x}-\mathbf{x}'|}\over |\mathbf{x}-\mathbf{x}'|}
\,\fn(-k^2)\right)\,,\label{Gretkn}
\eea
where $k=|\mathbf{k}|$, and $\fn(-k^2)\equiv\left.\Gkn(\rhB)\right|_{q^2=
-\mathbf{k} ^2}$.

The final integral in (\ref{Gretkn}) can be performed by a careful choice
of contour, subject to convergence of the integrand, which we have checked
numerically. It is found that for $n=0$, $\gg\Z0(-k^2)$ has a second
order pole at $k=0$, together with first order poles at $k=\pm iq\Z{0,j}$,
where $q\Z{0,j}>0$, $j=1,\dots,\infty$. For $n\ne0$, all poles occur at
$k=\pm iq\Z{n,j}$, where $q\Z{n,j}>0$, $j=1,\dots,\infty$. We close the
contour in the upper half plane, but perform a cut on the Im$(k)$ axis
on the interval $k\in(\half i\bar q,\infty)$, where $\bar q=\;\hbox{inf}\;
\{q_{n,j}| j=1,\dots,\infty\}$. Integrating back and forth around the cut,
first from $k=\eps+i\infty$ to $k=\eps+\half i\bar q$ in the Re$(k)>0$
quadrant and then back from $k=-\eps+\half i\bar q$ to $k=-\eps+i\infty$
in the Re$(k)<0$ quadrant before taking the limit $\eps\to0$,
has the net effect of circumscribing each of the poles on the positive
imaginary axis once in a clockwise fashion.

We will not analytically determine each coefficient in the sum of terms
in (\ref{Gretkn}) which result from the enclosed poles at $k=iq\Z{n,j}$, but
simply note that the Laurent expansion of $\fn(-k^2)$ at each of these poles
takes the form
\beq
\fn(-k^2)= {\cnj(\rhB)\over
k^2+q_{n,j}^2} + {\rm O}(1)\,,
\label{laurentj}
\eeq
in terms of coefficients $\cnj(\rhB)$, and the residue gives a Yukawa
correction in each case.

The pole at $k=0$ is not enclosed by the contour, but since it lies on
the contour, taking the principal part gives a net contribution to the
static potential, which is readily determined analytically by applying
identities which hold for the Legendre function solutions (\ref{exact0})
for $n=0$. In particular,
\bea \hbox{Res}\;\left(k\,\e^{ik|\mathbf{x}-\mathbf{x}'|}\gg\Z0(-k^2)
\right)_{k=0}&=&\hbox{Res}\;\left(-k\e^{ik|\mathbf{x}-\mathbf{x}'|}
\PP_{\nu-1}(y)\over\half\ga(1-y^2)\pt_y\PP_{\nu-1}(y)\right)_{k=0,\rh=\rhb}
\nonumber\\ &=&\hbox{Res}\;\left(-2k\e^{ik|\mathbf{x}-\mathbf{x}'|}
\PP_{\nu-1}(y)\over\ga\nu\left[y\PP_{\nu-1}(y)-P_\nu(y)\right]
\right)_{k=0,\rh=\rhb}\nonumber\\
&=&\lim_{\nu\to1}\left.4\LA\PP_{\nu-1}(y)\over\ga\left[y\pt_\nu\PP_{\nu-1}(y)-
\pt_\nu P_\nu(y)\right]\right|_{\rh=\rhb}\nonumber\\
&=&{2\LA(1+\rhB)\over\ga\rhB}\,,
\eea
where in the intermediate steps $y$ is defined implicitly by $y=(1-\rh)/
(1+\rh)$, and we have used the fact that as $k\to0$, $\nu\simeq1-k^2/(2\LA)$,
$\PP\Z0(y) =1$, and the identities $\lim_{\nu\to1}\pt_\nu\PP_{\nu-1}(y)=\ln[
\half(1+y)]$, and $\lim_{\nu\to1}\pt_{\nu} \PP_\nu(y)=y\ln[\half(1+y)]+y-1$.

The final expression for the static potential then becomes\footnote{Eq.\
(\ref{statpot}) corrects a small numerical factor in the Newton--like
term given in ref.\ \cite{benalex}.}
\beq
V_\Phi(\mathbf{x},\phi;\mathbf{x}',\phi')=
{-(1+\rhB)\LA\over4\pi^2\ga\rhB|\mathbf{x}-\mathbf{x}'|}
-\sum_{n=-\infty}^{\infty}\sum_{j=1}^{\infty} {\cnj(\rhB) e^{-q_{n,j}|
\mathbf{x}-\mathbf{x}'|}\over8\pi^2|\mathbf{x}-\mathbf{x}'|}\,.
\label{statpot}\eeq
which as expected is a Newtonian--type potential supplemented by
Yukawa--type corrections. The constant $\ga$ may be re-expressed in
terms of $\rM$ and $\rhB$ on account of (\ref{bega}).
\subsection{Nonlinear gravitational waves on the brane\label{branewave}}

We will now further justify the claim that the calculation of the
static potential for a minimally coupled massless scalar field given above
reproduces all the essential features of the corresponding calculation for
the graviton. We observe that the construction of nonlinear gravitational
waves which was developed in ref.\ \cite{LW} by a generalization of the
technique of Garfinkle and Vachaspati \cite{GV}, is unchanged when applied
to the background (\ref{thickbulkG}).
In particular, nonlinear gravitational waves can be constructed
on the background in the case that the geometry, $\Mlow$, generated
by the 4--dimensional metric $\gh_{\mu\nu}$ admits a hypersurface--orthogonal
null Killing vector, $k^\mu$. If $\bar\fu$ is a locally defined
scalar such that $\pt_{[\mu}k_{\nu]}=k_{[\nu}\pt_{\mu]}\bar\fu$ and
$k^\mu\pt_\mu\fu=0$, where Greek indices are lowered and raised with
$\gh_{\mu\nu}$ and its inverse, then the nonlinear wave spacetime is given
by adding to (\ref{thickbulkG}) the term
\beq
r^2H\e^{-\fu}k_\mu k_\nu\dd x^\mu\dd x^\nu,
\label{waveterm}\eeq
where $\fu$ is the pullback of $\bar\fu$ to (\ref{thickbulkG}), and $H$
is a scalar function on the bulk spacetime (\ref{thickbulkG}), which
satisfies $\nabla^a\nabla_a H=0$, and $\ell^a\pt_a H=0$. Here $\nabla_a$
is the covariant derivative in the metric (\ref{thickbulkG}) and $\ell^a
=(k^\mu,0,0)$ is the extension of $k^\mu$ to (\ref{thickbulkG}), with indices
raised and lowered by the full spacetime metric (\ref{thickbulkG}). The
vector, $\ell^a$, is also null and hypersurface orthogonal, and satisfies
$\pt_{[a}\ell_{b]}=\ell_{[b}\pt_{a]}(\fu+2\ln r)$ and $\ell^a\pt_a
(\fu+2\ln r)=0$.

In addition to the junction conditions (\ref{jc2}), (\ref{JC1})--(\ref{JCde}),
we now have the additional relation
\beq
{\sqrt{\DE(\rB)}\over\rB}\left.\pt_r(r^2H)\right|_\rb=\dsp{\eps T\over2}\,
\e^{-\la\ka\ph(\rb)}\rB^2 H(\rB)
\label{jcH}
\eeq
Using (\ref{JC2}) we see that (\ref{jcH}) is equivalent to the Neumann
condition
\beq
\left.\pt_rH\right|_\rb=0,
\eeq
at the brane, if $H$ is viewed as a massless scalar field on the spacetime
without the term (\ref{waveterm}). In the case that $\gh_{\mu\nu}=\eta
_{\mu\nu}$ the field $H$ therefore satisfies the same wave equation and
boundary conditions as were given above for massless scalar field, $\Phi$.

To make the correspondence explicit, we take $\gh_{\mu\nu}=\eta_{\mu\nu}$,
and adopt double null coordinates, $(u,v,x_\perp^1,x_\perp^2)$,
on the Minkowski space, $\Mlow$. If we choose
$k^\mu =(\partial_v)^\mu$, the solution with the gravitational wave term
(\ref{waveterm}) reads
\beq
\dd s^2=r^2\left[-\dd u\dd v +H(u, x_\perp^k,r,\ph)\dd u^2
+\de\Z{AB}\dd x_\perp^A \dd x_\perp^B \right]
+{r^2\dd r^2\over\DE}+\DE\dd\ph^2
\,,
\label{wave-on-eta}
\eeq
where $\DE$ is given by (\ref{trianglevar}).
Note that $H$ does not depend on $v$ but its dependence on $u$ is
arbitrary. The scalar wave equation for $H$ explicitly reads
\beq
H,_{rr} + \left({3\over r}+{\DE,_r\over\DE}\right)H,_r+{r^2H,_{\ph\ph}
\over\DE^2}+{\delta^{AB}H,_{AB}\over\DE}=0.
\label{Hcondition}
\eeq

The general linearized limit of the nonlinear gravitational solution can be
discussed as in~\cite{CG}. We note that
$H = h_{AB}(u) x_\perp^A x_\perp^B$, where $h\Z{22}(u)=-h\Z{11}(u)$, is clearly
a solution: it satisfies $\delta^{AB}H,_{AB}=0$, and its linearized limit is
analogous to the famous normalizable massless mode in the Randall--Sundrum
II model \cite{RS2}.
If we transform the radial parameter to $\rh$ by (\ref{rhodef})
and make a Fourier decomposition as in (\ref{Gphi}), then
(\ref{Hcondition}) becomes equivalent to the homogeneous part of
(\ref{sl}). Our analysis above for the massless scalar field
therefore applies equally to the graviton mode.

The nonlinear gravitational wave construction also applies
firstly to any other Ricci--flat geometry on $\Mlow$ which admits a
hypersurface orthogonal null Killing vector, and secondly with suitable
modifications to other Einstein space geometries for $\Mlow$, provided
appropriate solutions can be found.

\section{The hierarchy problem\label{hierarchy}}

One of the principal motivations for studying brane world models is
the attempt to provide a natural solution to the hierarchy problem
between the Planck and electroweak scales.
The construction of ref.\ \cite{LW} potentially offers a concrete realization
of the phenomenological solution to the hierarchy problem proposed by
Antoniadis \cite{Anto}, Arkani-Hamed, Dimopoulos
and Dvali \cite{ADD,AADD}. In particular, if the non-gravitational
forces can be introduced in such a way as to be confined to the brane, then
provided that the distance between the thin brane and the bolt can be
made large enough, higher--dimensional gravitational corrections could
become manifest close to the TeV scale.

Since the construction is a hybrid one, there is an ordinary \KK\ direction
within the 4--brane in addition to the direction transverse to the brane.
A phenomenologically realistic solution to the hierarchy problem can
therefore only be obtained if the distance between the brane and the bolt
can be made many orders of magnitude larger than the circumference of the
\KK\ circle. In the original construction of ref.\ \cite{LW}, based on
Einstein--Maxwell gravity with a higher--dimensional cosmological constant,
a natural solution to the hierarchy problem proved to be impossible as the
brane--bolt distance was at most comparable to the circumference of the
\KK\ circle. The present model has more degrees of freedom, however, and
so it is possible that this problem can be overcome.

In order to make the volume of the internal
space, $\VV$, sufficiently large to accommodate TeV scale gravity, we
must be able to find a set of parameters $(A,B,\LA)$ which allows
the ratio, $R=\VV/\CC$, to be arbitrarily large.
The proper circumference of the \KK\ direction is
\beq
\CC
={4\pi\sqrt{\DE(\rB)}\over\DE'(\rM)}
={2^{3/4}\pi\rhB^{1/2}\over\sqrt{\LA}\,\sqrt[4]{1+\rhB}}\,.
\label{KKcirc}\eeq
The volume of the internal space,
$\VV=2\int_{r\X{\!-}}^{\rb}\int\Z0^{4\pi/\DE'(r\X{\!-})}
\dd\ell_{\th}\dd\ell_{r}$ is
\beq
\VV
=({\rB}^2-{\rM}^2){4\pi\over\DE'(\rM)}
\eeq
The ratio, $R=\VV/\CC$ is then simply
\beq
R={({\rB}^2-{\rM}^2)\over\sqrt{\DE(\rB)}}
=\sqrt{2\over\LA}\,\rM\,F(\rhB)
\label{rat1}\eeq
where $\rM$ is given by (\ref{rplusminus}),
\beq
F(\rhB)={4\rhB^{1/2}\over1-\rhB}\left(\sqrt{2\over1+\rhB}-1\right)\sqrt[4]
{1+\rhB\over2}\,,
\label{rat2}\eeq
and on account of (\ref{rplusminus}) and (\ref{rho_b}),
\beq
\rhB={A-\sqrt{A-\half B^2\LA}\over A+\sqrt{A-\half B^2\LA}}\,.
\eeq

\EPSFIGURE[p]
{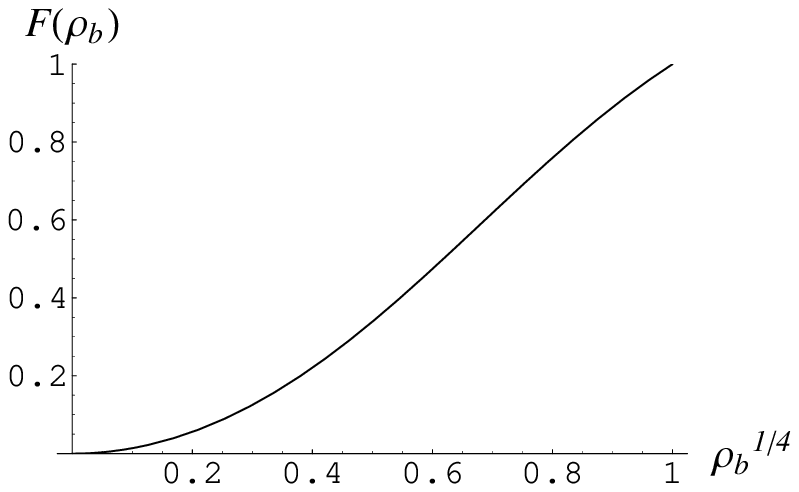}
{The ratio of volume of the internal space to its circumference is a
multiple of $F(\rhB)$, as given by (\ref{rat1}), (\ref{rat2}). It is plotted
here versus $\rhB^{1/4}$.
\label{fz}}
The quantity $F(\rhB)$ defined by (\ref{rat2}) is depicted in Fig.~\ref{fz}.
It is a monotonic function which increases from $F=0$ to $F=1$ on the
interval $\rhB\in[0,1]$. The limit $\rhB\to1$ occurs when ${B^2\LA\over 2A^2}
\to1$, i.e., when $\rM\approx \rP\approx \rB\approx {4A\over\LA }$.
In this case $\CC\approx\sqrt{2}\pi\LA^{-1/2}$, and $R=\VV/\CC\approx
2^{5/2}A\LA^{-3/2}$. The requirement that $\CC$ must be small enough
to be interpreted as a conventional \KK\ direction means that $\LA$ must
be suitably large. Since the parameter $A$ is still free, however, we
can still make $R$ arbitrarily large to overcome its dependence on the
$\LA^{-3/2}$ factor. Thus it appears that a solution to the hierarchy
problem may be feasible.

For smaller values of $\rhB$ similar arguments apply. In particular,
consider the extreme limit $\rhB\to0$ which corresponds to $0<\frac{B^2\LA}
{2A^2}\ll1$. Then $\rM^4\approx B^2/A$, $\rP^4\approx8 A/\LA$ and $F(\rhB)
\approx4(\sqrt{2}\,-1)\rhB^{1/2}$. Hence $R\approx2(\sqrt{2}\,-1)B^{3/2}
A^{-5/4}$ and $\CC\approx 2^{-3/4}\pi BA^{-1}$, which are both independent
of $\LA$. Since the constants $A$ and $B$ are not constrained except by the
requirement $\frac BA\ll\sqrt{\frac2{\LA}}$, we can again make $R$
arbitrarily large while keeping $\CC$ small. If we denote $R\Z0$ and $\CC\Z0$
to be phenomenologically desirable values of $R$ and $\CC$,
we can conversely fix both $A$ and $B$. We find $A_0=\pi^6{R\Z0}^4\,
2^{-17/2}(\sqrt{2}\,-1)^{-4}\CC\Z0^{-6}$ and $B_0=\pi^5{R\Z0}^4\,2^{-31/4}
(\sqrt{2}\,-1)^{-4}\CC\Z0^{-5}$, while for consistency of the limit
$0<\LA_0\ll\frac{\pi^2}{\sqrt{2}{\CC\X0}^2}$.

We have demonstrated here that a solution of the hierarchy problem is
possible regardless of the value of $\rhB$. We note, however, that once the
Newton potential for the tensor modes, equivalent to the first term of
(\ref{statpot}), is determined then two of the parameters, $r_-$, $\LA$ and
$\rhB$ would be fixed phenomenologically via equations similar to
(\ref{statpot}) and (\ref{KKcirc}), in terms of the Newton constant and the
energy scale for the ordinary \KK\ circle direction. From (\ref{rat1})
we see that just enough parameter freedom remains to choose the remaining
independent parameter to solve the hierarchy problem as desired.

\section{Conclusion}

We have extended the construction of ref.\ \cite{LW} to produce a new
hybrid brane world compactification in six dimensions with a number of
desirable features. As is the case with the earlier model, the observable
universe corresponds to a codimension one brane which
has one extra \KK\ direction and which closes regularly in the bulk
at bolts, namely geodesically complete submanifolds where a rotational
Killing vector $\pt/\pt\th$ vanishes. The regularity of the geometry
ensures that construction avoids potential problems that often arise
when extra matter is added to models with additional horizons or
singularities in the bulk. The construction of nonlinear gravitational
waves in \S4.4 is an explicit demonstration of this. Furthermore,
we have demonstrated that such gravitational wave equations include a mode
which may be considered as a massless minimally coupled scalar field
on the unperturbed bulk geometry, with Neumann boundary conditions at the
brane, and that such a mode has a static potential with a long
range Newtonian potential plus Yukawa corrections. As discussed in
\S\ref{sec:morehybrid} further non--trivial calculations are required
to determine the additional gauge field content and associated spectrum of
Kaluza--Klein excitations resulting from the Kaluza--Klein circle
contribution to the dimensional reduction.

The most significant improvement that the present model has over the
earlier construction of ref.\ \cite{LW}, is that the supersymmetric
Salam-Sezgin action allows a hybrid brane world construction in which there
appears to be just enough parameter freedom to make a solution to the
hierarchy problem feasible. For those parameter ranges which achieve
this, giving a deep bulk direction as compared to the radius of the
\KK\ circle, it is quite possible that the spacing of the Yukawa levels
would become so close that their sum would approximate inverse powers of
$|\mathbf{x}-\mathbf{x}'|$ rather than a single Yukawa--like term.
Such corrections would then be similar to those of the Randall--Sundrum
II model \cite{RS2}.

In comparison to brane world models in six dimensions which view the
physical universe as a codimension two topological defect, we note
that the position of the four-brane in the bulk is uniquely determined by
the bulk geometry and does therefore not require the addition of other
branes in the bulk, or of special matter field configurations on them. The
degree of naturalness by which the cosmological constant problem might be
solved in this model is an interesting open problem which we have not pursued.

In order to solve the field equations analytically it was necessary to assume
that the 4--dimensional cosmological constant was zero. However, our
construction does not seem to preclude the possibility of the model having a
non-zero cosmological constant in four dimensions, similar to the explicit
solutions found for the model of \cite{LW}. The analysis of Appendix
\ref{sec:bh} suggests that such solutions exist but are unlikely to have
a simple analytic form. Even though they would be non--singular in the
bulk, the existence of such solutions in not precluded by the recent
uniqueness theorem of Gibbons, G\"uven and Pope \cite{GGP}, since the
presence of the codimension one brane provides a loophole to its proof.
If analytic solutions with non--zero 4--dimensional cosmological constant
could be found, then the nonlinear gravitational wave construction of
\S\ref{branewave} should generalize directly. Examples of the bulk solutions
in question have recently been given numerically by Tolley {\it et al.}\
\cite{TBHA}. It would also be interesting to consider
the influence that matter sources on the brane would have on such
solutions, a question that has recently been considered at the linearised
level in other 6--dimensional models \cite{dRT}.

Even in the absence of a cosmological constant, the solutions
(\ref{thickbraneG}), (\ref{thickF}), (\ref{fofr})--(\ref{trianglevar}),
together with the hybrid construction offer the possibility of
generating brane world black hole solutions as well as the gravitational
wave solutions already presented. Since the solutions given
apply to arbitrary Ricci--flat geometries in the physical 4--dimensions,
they include the Schwarzschild and Kerr geometries as particular examples.
The most important open problem is an analysis of gravitational perturbations
on such backgrounds analogous to the case of the 4--dimensional flat
background studied in \S{\ref{sec:Potential}. Such an analysis would
resolve the important question of the stability of such black holes in
the 6--dimensional setting, and also give some idea of potential signatures
of higher dimensions on black hole physics. Given that the construction
of brane world black holes is generally far from trivial, the hybrid
compactifications offer a promising arena for studying concrete
realizations of such solutions.

In conclusion, we believe that the construction of ref.\ \cite{LW}
combines some of the best features of both the Randall--Sundrum and
\KK\ scenarios, and leads naturally to a class of hybrid compactifications
which should be further studied. The present paper shows that the
extension to bulk geometries of the supersymmetric Salam--Sezgin background
provides further phenomenological reasons for doing so.

\acknowledgments
We thank Jorma Louko for valuable discussions. This work was supported by the
Marsden Fund of the Royal Society of New Zealand.

\appendix
\section{General fluxbrane and dual static black hole-like solutions}
\label{sec:bh}

The global properties of certain static solutions of electrically charged
dilaton spacetimes with a dilaton potential of Liouville form were classified
in ref.\ \cite{PW} without explicitly writing down the general solution.
The solutions considered in ref.\ \cite{PW} include spherically symmetric
spacetimes, but in the most general case include geometries for which the
spatial sections at spatial infinity consist of an arbitrary Einstein space,
rather than simply a $(D-2)$-sphere in the case of $D$ spacetime
dimensions. Electrically charged solutions with these symmetries are of
interest, since in cases in which a regular horizon exists fluxbranes may
be obtained from them by the double analytic continuation technique
that was first introduced in \cite{GW}. The field equations
considered in ref.\ \cite{PW} include our equations (\ref{Fe1})--(\ref{Fe1})
as a special case.

At a first glance, it would appear that the Salam-Sezgin model is a special
case of the class of models analysed in ref.\ \cite{PW}. Unfortunately,
however, the particular coupling constants which appear in the exponential
coupling of the scalar to the $U(1)$ gauge field, and the Liouville potential,
are in fact a degenerate case of the analysis of ref.\ \cite{PW}.

In this Appendix we will therefore repeat the analysis of \cite{PW}
in the case of the Salam-Sezgin model, but in a slightly more general
framework which incorporates fluxbranes at the outset, in addition to
their dual solutions. Rather than simply restricting our
attention to the Salam-Sezgin model in six dimensions, we will investigate
relevant solutions for the whole degenerate case omitted in \cite{PW}.
The relevant field equations are those which follow from variation
of the $D$-dimensional action\goodbreak
\bea
S=\int\dd^Dx\sqrt{-g}\Bigl\{{{\cal R}\over4\ka^2}&-&{1\over\Dm2}\,g^{ab}\pt_a
\ph\,\pt_b\ph-{1\over4}\exp\left(4\ka\ph\over\Dm2\right)F_{ab}F^{ab}
\nonumber\\ &-&{\LA\over2\ka^2}\exp\left(-4\ka\ph\over\Dm2\right)\Bigr\},
\label{action_D}
\eea
The field content of (\ref{action_D}) is the same as that of action
(\ref{action}) in the absence of the Kalb-Ramond field, and the field
equations obtained by variation of this action reduce to
(\ref{Fe1})--(\ref{Fe1}) (\ref{action}) when $D=6$. The model of ref.\
\cite{PW} was more general than (\ref{action_D}) in allowing
for two additional arbitrary coupling constants: one in the dilaton /
$U(1)$ coupling, and one in the Liouville potential. In the notation
of ref.\ \cite{PW} our conventions are the same, but we have chosen
$\g0=-1$ and $\g1=1$: in this case the results of \cite{PW} are degenerate.

The field equations obtained by varying (\ref{action_D}) are most easily
integrated explicitly in the static case by using the radial coordinate
of Gibbons and Maeda \cite{GM}, for which the metric is given by
\beq
\dd s^2=\ee\e^{2u}\left[-\eqq\dd t^2+R^{2\Db2}\dd\xi^2\right]+R^2\bar g_{ij}
\dd\xt^i\dd\xt^j,\label{coordb}
\eeq
where $u=u(\xi)$, $R=R(\xi)$, and $\bar g_{ij}$ is the metric of a
$\Db2$-dimensional Einstein space,
\beq
\bar{\cal R}_{ab}=\Db3\lb\,\bar g_{ab},\qquad a,b=1,\dots,\Dm2,
\eeq
$\eqq=\pm1$ and $\ee=\pm1$. If $\eqq=+1$ and $\ee=+1$ one obtains the
geometry relevant to the domain of outer communications of a black
hole, or of a naked singularity. The case $\eqq=+1$ and $\ee=-1$ would
correspond to the interior of a black hole in the case that regular
horizons exist. If we take $\eqq=-1$ and $\ee=+1$ we have the case of
a fluxbrane, assuming $t$ to be an angular coordinate.

We choose $\bf F$ to be the field of an isolated electric charge,
\beq
{\bf F}=\exp\left(2u-{4\ka\ph\over\Dm2}\right){Q\over\ka}\;\dd t
\wedge\dd\xi\,,\label{Fb}
\eeq
in the case that $\eqq=+1$, and a magnetic field
in the case that $\eqq=-1$. In the later case the ansatz (\ref{Fb})
is the same, except that $t$ is now an angular coordinate and $Q$ is
the magnetic charge.

With the ansatz (\ref{coordb}), (\ref{Fb}) and assuming $\ph=\ph(\xi)$,
the field equations can be written \cite{PW} as the system
\bea
\ddot\et&=&2\ee\ekQ\label{junka}\\
\ddot\ze&=&\Db3^2\ee\elb-2\ee\eLA,\label{junkb}\\
\ddot\ch&=&\Db2\Db3\ee\elb-2\ee\eLA,\label{junkc}
\eea
with the constraint
\beq
\Db2\left[\dot\ze^2-2\dot\ze\dot\ch\right]+\Db3\dot\ch^2+
\dot\et^2+\Db2\Db3\ee\elb-2\ee\eLA-2\ee\ekQ=0,
\label{constraint}
\eeq
where the overdot denotes $\dd/\dd\xi$. Eq.\ (\ref{junka}) is readily
integrated if we multiply it by $\dot\et$, yielding
\beq
\dot\et^2=2\ekQ+\ep\Z2\Db2{k\Z2}^2,
\eeq
where $k\Z2$ is an arbitrary constant and $\ep\Z2=+1,0,-1$. If
$\eqq=+1$ (``black hole'' case) then a further integration yields three
possible solutions, distinguished by the parameter $\ep\Z2$:
\beq
{2Q^2\over\Dm2}e^{2\et}=\cases{{k\X2^{\ 2}\over\sinh^2\left[\sqrt{\Dm2}\;k\X2
\xixi2\right]},&$\ep\Z2=+1$,\cr {1\over\left(\Dm2\right)\xixi2^2},&$\ep\Z2=\ 0$
,\cr{k\X2^{\ 2}\over\sin^2\left[\sqrt{\Dm2}\;k\X2\xixi2\right]},&$\ep\Z2=-1$,
\cr}
\eeq
where $\xi\Z2$ is an arbitrary constant. If $\eqq=-1$ (``fluxbrane''
case) then we must have $\ep\Z2=+1$ and the only solution is
\beq
{2Q^2\over\Dm2}e^{2\et}={k\Z2^{\ 2}\over\cosh^2\left[\sqrt{\Dm2}\;k\X2
\xixi2\right]}\,.
\eeq

Linear combinations of (\ref{junkb}) and (\ref{junkc}) yield
\bea
\ddot\ch-\ddot\ze&=&\Db3\ee\elb,\label{junkd}\\
\Db3\ddot\ch-\Db2\ddot\ze&=&2\ee\eLA,\label{junke}
\eea
while the constraint (\ref{constraint}) becomes
\beq
\dot\ze^2-2\dot\ze\dot\ch+\left(D-3\over D-2\right)\dot\ch^2+
\Db3\ee\elb-{2\ee\eLA\over D-2}+\ep\Z2{k\Z2}^2=0.
\label{constraint1}
\eeq

In the special cases that $\LA=0$, or $\lb=0$, eqs.\
(\ref{junke})--(\ref{constraint1}), can be further integrated, as follows:

\noi {\bf(i)} Special $\LA=0$ solution (as previously given in \cite{PW}):
\beq
\ee\elb=\cases{{k\X1^{\ 2}\over\sinh^2\left[(\Dm3)k\X1\xixi1\right]},&
$\ep\Z1=+1$, $\ee\lb>0$,\cr {1\over(\Dm3)^2\xixi1^2},&$\ep\Z1=\ 0$,
$\ee\lb>0$,\cr{k\X1^{\ 2}\over\sin^2\left[(\Dm3)k\X1\xixi1\right]},&
$\ep\Z1=-1$, $\ee\lb>0$,\cr {-k\X1^{\ 2}\over\cosh^2\left[(\Dm3)
k\X1\xixi1\right]},& $\ep\Z1=+1$, $\ee\lb<0$,\cr}
\label{solsp}\eeq
where $\xi\Z1$ is and arbitrary constant and
\beq
\Db3\ep\Z1{k\Z1}^2=\ep\Z2\Db2{k\Z2}^2+\left(D-3\over D-2\right){c\Z1}^2,
\label{con1}\eeq
with $k\Z1$, $c\Z1$ constants constrained only by the requirement that
(\ref{con1}) have real solutions.

\noi {\bf(ii)} Special $\lb=0$ solution:
\beq
2\eeLA\e^{2\ch}=\cases{{-k\X3^{\ 2}\over\sinh^2\left[k\X3\xixi3
\right]},& $\ep\Z3=+1$, $\eeLA<0$,\cr {-1\over\xixi3^2},
&$\ep\Z3=\ 0$, $\eeLA<0$,\cr{-k\X3^{\ 2}\over\sin^2\left[k\X3
\xixi3\right]},& $\ep\Z3=-1$, $\eeLA<0$,\cr {k\X3^{\ 2}\over\cosh^2
\left[k\X3\xixi3\right]},& $\ep\Z3=+1$, $\eeLA>0$,\cr}
\eeq
where $\xi\Z3$ is and arbitrary constant and
\beq
{\ep\Z3{k\Z3}^2\over D-2}=\ep\Z2{k\Z2}^2+{c\Z3}^2,
\label{con2}\eeq
with $k\Z3$, $c\Z3$ constants constrained only by the requirement that
(\ref{con2}) have real solutions. The solution for the $\lb=0$ Salam--Sezgin
fluxbrane ($D=6$, $\eqq=-1$, $\ep\Z2=+1$, $\ee>0$, $\LA>0$) has been given
previously in terms of these variables by Gibbons, G\"uven and Pope \cite{GGP},
and is readily seen to agree with the above upon making the replacements
$\et\to x$, $\ch\to y$, $2(\ze-\ch)\to z$, $k\Z2\to\half\la\Z1$, $k\Z3\to
\la\Z2$, $c\Z2\to\half\la\Z3$, to make contact with their notation.

The general solution other than in the special cases
(\ref{solsp})--(\ref{con2}) does not appear to have an obvious simple
analytic form. However, general properties of the solutions can be gleaned
following the method of \cite{PW}. The constraint (\ref{constraint1}) may
be used to eliminate $\ee\elb$ from (\ref{junkd}), to yield a 3--dimensional
autonomous system of first--order ODEs. If we define $X\equiv\dot\ze$,
$V\equiv\dot\ch$ and $W\equiv\sqrt{2}\,\e^\ch/\sqrt{D-2}$, this system is
given by
\bea
\dot X&=&-\Db3P-\ee\LA W^2\\ \dot V&=&-\Db2P\\ \dot W&=&VW
\eea
where
\beq
P\equiv X^2-2XV+\left(D-3\over D-2\right)V^2+\ep\Z2{k\Z2}^2.
\label{constraint2}\eeq
The fact we have a 3--dimensional system means that the analysis is
considerably simpler than in the 5--dimensional examples of ref.\ \cite{PW},
and is closer to the phase space of a simple spherically symmetric
uncharged black hole with a Liouville potential \cite{MW}.

Trajectories with $W=0$ remain confined to the plane. Consequently, in the
full 3--dimensional phase space we can take $W\ge0$ without loss of
generality.

As is the case in refs. \cite{PW,MW} the only critical points at a finite
distance from the origin are given by the 1--parameter locus of points
with $W=0$ and $P=0$. From (\ref{constraint2}) it follows that the critical
points are: (i) hyperbolae in the first and third quadrants of the $W=0$ plane
if $\ep\Z2>0$; (ii) straight lines $V=\sqrt{D-2}\left[\sqrt{D-2}\pm1\right]X
/(D-3)$ if $\ep\Z2=0$; and (iii) hyperbolae which cross all quadrants if
$\ep\Z2<0$. The $W=0$, $P=0$ curve is described by the locus $(X\Z0,V\Z0)$,
where
\beq
V\Z0={\sqrt{D-2}\over D-3}\left[\sqrt{D-2}\,X\Z0\pm
\sqrt{{X\Z0}^2-\left(D-3\right)\ep\Z2{k\Z2}^2}\right]\,.
\label{crit}\eeq
These critical points are found to correspond to asymptotic regions
$R\to\infty$, or singularities with $R\to0$, except in
the special case that $X\Z0=V\Z0=\sqrt{D-2}\,k\Z2$. For the special
case $R\to\const$, which represents a horizon in the black hole case,
or a bolt in the fluxbrane case.

The integral curves which lie in the plane $W=0$ are the lines
\beq V=\left(D-2\over D-3\right)X+\const,\label{intlines}\eeq
and these of course correspond to the special solutions
(\ref{solsp}), (\ref{con1}). Such lines cross the $W=0$, $P=0$ curve once
in the first quadrant and once in the third quadrant if $\ep\Z2=0,+1$.

An analysis of small perturbations about the critical points (\ref{crit})
shows that the eigenvalues are $\{0,2X\Z0,\,V\Z0\}$.
Thus points in the first quadrant repel a 2-dimensional bunch of
trajectories out of the $W=0$ plane, while points in the third quadrant attract
a 2-dimensional bunch of trajectories out of the $W=0$ plane for all
values of $\ep\Z2$. For $\ep\Z2=-1$ the points in the second and fourth
quadrants are saddle points with respect to directions out of the $W=0$
plane. Points in the second quadrant each attract one of the lines
(\ref{intlines}) in $W=0$ plane, while points in the fourth quadrants
similarly each eject one of the lines (\ref{intlines}).

We will henceforth restrict our attention to the case $\ee=+1$, so that
we are dealing with the domain of outer communications in the case of a
black hole or naked singularity ($\eqq=+1$); or with a fluxbrane ($\eqq=-1$).

The following critical points are found at the phase space infinity, and
coincide with a subset of the critical points of the more general system
of ref.\ \cite{PW}. We will label them identically to the notation of
ref.\ \cite{PW}. The points are:
\begin{itemize}
\item $L\Z{5-8}$ located at $X=\pm\infty$, $V=[D-2\pm\sqrt{D-2}]X/(D-3)$,
$W=0$. These points are the endpoints of the 1--parameter family of critical
points with $P=0$ at finite values of $X$ and $V$ in the $W=0$ plane.
The eigenvalues for small perturbations are again $\{0,2X\Z0,\,V\Z0\}$.
\item $M\Z{1,2}$ located at $X=\pm\infty$, $V=\left(D-2\over D-3\right)X$,
$W=0$. These points correspond to asymptotically flat solutions, and have
$P=-X^2/(D-3)$ and $\lb>0$. The eigenvalues for small perturbations are
$\{-1,-1,1/(D-3)\}$. The two dimensional set of solutions attracted are
simply the integral curves (\ref{intlines}) which represent solutions for
the system with no scalar potential, i.e., $\LA=0$.
\item $P\Z{1,2}$ located at $X=\pm\infty$, $V=X$, $W=|X|/\sqrt{-
(D-2)\LA}$. These points only exist if $\LA<0$, and have
$P=-X^2/(D-2)$ and $\lb=0$. They are thus are endpoints for
integral curves with all possible signs of $\lb$. The eigenvalues for small
perturbations are $\{-1,-1,0\}$. It is quite possible that higher order
terms would lift the degeneracy of the zero eigenvalue. However, we will
not investigate this further, as the solutions with $\LA<0$ are not our
prime concern in this paper. The points $P\Z{1,2}$ represent the $r\to\infty$
asymptotic region for solutions which are not asymptotically flat, but
which have the unusual asymptotics listed in Table II of ref.\ \cite{PW}
with $g\Z1=-1$.
\end{itemize}

Although the present model is a degenerate case of the more general
analysis of ref.\ \cite{PW}, all of the possible asymptotic properties
of the solutions outlined above are special cases of the analysis of
\cite{PW}, and thus the general conclusions obtained there also hold
here. In particular, there are no regular black hole solutions with $\lb>0$
apart from a class with unusual asymptotics which exist if $\LA<0$. In the
case of the Salam--Sezgin model, $\LA>0$, and so no regular uncompactified
black holes exist in that case.

For the purposes of the construction of \cite{LW}, which we
generalize in this paper, the particulars of the asymptotic solutions are
not important, however, since part of the spacetime is excluded once the
thin brane is inserted in the fluxbrane background. Whether or not dual
black holes with standard (or even unusual) asymptotic properties exist
is therefore not of primary importance. What is important is that the
spacetime from the bolt to the thin brane should be regular. Provided
a regular horizon exists in the black hole case, which is dual to a bolt
in the fluxbrane, then the construction of ref.\ \cite{LW} should lead
to regular hybrid compactifications. The analysis above shows that such
solutions can be obtained only in the special case that $X\Z0=V\Z0=
\sqrt{D-2}\,k\Z2$, as there then exists a 2-dimensional bunch of
trajectories with any sign of $\lb$, including the $\lb>0$ case relevant
to a positive cosmological term on the brane.

We therefore believe that the construction used in this paper can be
extended to a small class of solutions with $\lb>0$ in the case of the
Salam--Sezgin model. Since it appears that such solutions could only be
constructed numerically \cite{TBHA}, we have not investigated them in further
detail. We have no reason to suspect that the qualitative properties of
the hybrid compactifications on such backgrounds should differ from those
of the $\lb=0$ solutions.


\begin{thebibliography}{66}

\bibitem{Ak}
K.~Akama,
{\it``Pregeometry''},
in K.~Kikkawa, N.~Nakanishi and H.~Nariai (eds),
{\it``Gauge Theory and Gravitation''},
{\it Lect.\ Notes Phys.\ } {\bf 176} (1982) 267 [arXiv:\hepth{0001113}].

\bibitem{RuSha} V.~A.~Rubakov and M.~E.~Shaposhnikov,
{\it``Do we live inside a domain wall?''},
\plb{125}{1983}{136}.

\bibitem{Vis} M.~Visser,
{\it``An exotic class of Kaluza-Klein models''},
\plb{159}{1985}{22} [arXiv:\hepth{9910093}];

\bibitem{GW} G.~W.~Gibbons and D.~L.~Wiltshire,
{\it``Spacetime as a membrane in higher dimensions''},
\npb{287}{1987}{717} [arXiv:\hepth{0109093}].

\bibitem{RS1}
L.~Randall and R.~Sundrum,
{\it``A large mass hierarchy from a small extra dimension''},
\prl{83}{1999}{3370} [arXiv:\hepph{9905221}].

\bibitem{RS2}
L.~Randall and R.~Sundrum,
{\it``An alternative to compactification''},
\prl{83}{1999}{4690} [arXiv:\hepth{9906064}].

\bibitem{CN}
Z.~Chacko and A.~E.~Nelson,
{\it``A solution to the hierarchy problem with an infinitely large
extra dimension and moduli stabilization''},
\prd{62}{2000}{085006} [arXiv:\hepth{9912186}].

\bibitem{CLP}
J.~W.~Chen, M.~A.~Luty and E.~Ponton,
{\it``A critical cosmological constant from millimeter extra
dimensions''},
\jhep{09}{2000}{012} [arXiv:\hepth{0003067}].

\bibitem{KMO}
P.~Kanti, R.~Madden and K.~A.~Olive,
{\it``A 6-D brane world model''},
\prd{64}{2001}{044021} [arXiv:\hepth{0104177}].

\bibitem{LMW}
F.~Leblond, R.~C.~Myers and D.~J.~Winters,
{\it``Consistency conditions for brane worlds in arbitrary
dimensions''},
\jhep{07}{2001}{031} [arXiv:\hepth{0106140}].

\bibitem{LW}
J.~Louko and D.~L.~Wiltshire,
{\it``Brane worlds with bolts''},
\jhep{02}{2002}{007} [arXiv:\hepth{0109099}].

\bibitem{BCCF}
C.~P.~Burgess, J.~M.~Cline, N.~R.~Constable and H.~Firouzjahi,
{\it``Dynamical stability of six-dimensional warped brane-worlds''},
\jhep{01}{2002}{014} [arXiv:\hepth{0112047}].

\bibitem{PK}
D.~K.~Park and H.~s.~Kim,
{\it``Single 3-brane brane-world in six dimension''},
\npb{650}{2003}{114} [arXiv:\hepth{0206002}].

\bibitem{ABPQsb}
Y.~Aghababaie, C.~P.~Burgess, S.~L.~Parameswaran and F.~Quevedo,
{\it``SUSY breaking and moduli stabilization from fluxes in gauged 6D
supergravity''},
\jhep{03}{2003}{032} [arXiv:\hepth{0212091}].

\bibitem{ABPQcc}
Y.~Aghababaie, C.~P.~Burgess, S.~L.~Parameswaran and F.~Quevedo,
{\it``Towards a naturally small cosmological constant from branes
in 6D supergravity''},
\npb{680}{2004}{389} [arXiv:\hepth{0304256}].

\bibitem{CDGV}
J.~M.~Cline, J.~Descheneau, M.~Giovannini and J.~Vinet,
{\it``Cosmology of codimension-two braneworlds''},
\jhep{06}{2003}{048} [arXiv:\hepth{0304147}].

\bibitem{BNQTZ}
C.~P.~Burgess, C.~Nunez, F.~Quevedo, G.~Tasinato and I.~Zavala,
{\it``General brane geometries from scalar potentials: Gauged
supergravities and accelerating universes''},
\jhep{08}{2003}{056} [arXiv:\hepth{0305211}].

\bibitem{ABCFPQTZ}
Y.~Aghababaie, C.~P.~Burgess, J.~M.~Cline, H.~Firouzjahi,
S.~Parameswaran, F.~Quevedo, G.~Tasinato and I.~Zavala,
{\it``Warped brane worlds in six dimensional supergravity''},
\jhep{09}{2003}{037} [arXiv:\hepth{0308064}].

\bibitem{Burgess}
C.~P.~Burgess,
{\it```Towards a natural theory of dark energy: Supersymmetric
large extra dimensions''},
AIP Conf.\ Proc.\ {\bf 743} (2005) 417 [arXiv:\hepth{0411140}];
{\it``Supersymmetric large extra dimensions and the cosmological
constant problem''},
arXiv:\hepth{0510123}.

\bibitem{Navarro}
I.~Navarro and J.~Santiago,
{\it``Gravity on codimension 2 brane worlds''},
\jhep{02}{2005}{007} [arXiv:\hepth{0411250}].

\bibitem{CV}
J.~Vinet and J.~M.~Cline,
{\it``Codimension-two branes in six-dimensional supergravity and the
cosmological constant problem''},
\prd{71}{2005}{064011} [arXiv:\hepth{0501098}].

\bibitem{LL}
H.~M.~Lee and C.~Ludeling,
{\it``The general warped solution with conical branes in six-dimensional
supergravity''},
\jhep{01}{2006}{062} [arXiv:\hepth{0510026}].

\bibitem{CB}
P.~Callin and C.~P.~Burgess,
{\it``Deviations from Newton's law in supersymmetric large extra
dimensions''},
arXiv:\hepph{0511216}.

\bibitem{TBHA}
A.~J.~Tolley, C.~P.~Burgess, D.~Hoover and Y.~Aghababaie,
{\it``Bulk singularities and the effective cosmological constant
for higher co-dimension branes''},
arXiv:\hepth{0512218}.

\bibitem{PST}
M.~Peloso, L.~Sorbo and G.~Tasinato,
{\it``Standard 4d gravity on a brane in six dimensional flux
compactifications''},
\prd{73}{2006}{104025} [arXiv:\hepth{0603026}].

\bibitem{HM}
G.~T.~Horowitz and R.~C.~Myers,
{\it``The AdS/CFT correspondence and a new positive energy conjecture
for general relativity''},
\prd{59}{1999}{026005} [arXiv:\hepth{9808079}].

\bibitem{MSYK}
S.~Mukohyama, Y.~Sendouda, H.~Yoshiguchi and S.~Kinoshita,
{\it``Warped flux compactification and brane gravity''},
\newjournal{J.\ Cosmol.\ Astropart.\ Phys.\ }{JCAPA}{07}{2005}{013}
[arXiv:\hepth{0506050}];
{\it``Dynamical stability of six-dimensional warped flux
compactification''},
arXiv:\hepth{0512212}.

\bibitem{NS}
H.~Nishino and E.~Sezgin,
{\it``Matter and gauge couplings of $N=2$ supergravity in six
dimensions''},
\plb{144}{1984}{187}.

\bibitem{SS}
A.~Salam and E.~Sezgin,
{\it``Chiral compactification on Minkowski $\times S^2$ of $N=2$
Einstein-Maxwell supergravity in six dimensions''},
\plb{147}{1984}{47}

\bibitem{GT}
R.~Geroch and J.~H.~Traschen,
{\it``Strings and other distributional sources in general
relativity''},
\prd{36}{1987}{1017}.
\bibitem{Ga}
D.~Garfinkle,
{\it``Metrics with distributional curvature''},
\cqg{16}{1999}{4101} [arXiv:\grqc{9906053}].

\bibitem{dRT}
C.~de Rham and A.~J.~Tolley,
{\it``Gravitational waves in a codimension two braneworld''},
arXiv:\hepth{0511138}.

\bibitem{KK}
N.~Kaloper and D.~Kiley,
{\it``Exact black holes and gravitational shockwaves on
codimension--2 branes''},
arXiv:\hepth{0601110}.

\bibitem{PW} S.~J. ~Poletti and D.~L.~Wiltshire,
{\it``The global properties of static spherically symmetric charged
dilaton spacetimes with a Liouville potential''},
\prd{50}{1994}{7260}
[Erratum-ibid.\ D {\bf52} (1995) 3753] [arXiv:\grqc{9407021}].

\bibitem{CHM}
K.~C.~K.~Chan, J.~H.~Horne and R.~B.~Mann,
{\it``Charged dilaton black holes with unusual asymptotics''},
\npb{447}{1995}{441} [arXiv:\grqc{9502042}].

\bibitem{GH}
G.~W.~Gibbons and S.~W.~Hawking,
{\it``Classification of gravitational instanton symmetries''},
\cmp{66}{1976}{291}.

\bibitem{GKR}
S.~B.~Giddings, E.~Katz and L.~Randall,
{\it``Linearized gravity in brane backgrounds''},
\jhep{03}{2000}{023} [arXiv:\hepth{0002091}].

\bibitem{GV}
D.~Garfinkle and T.~Vachaspati,
{\it``Cosmic string traveling waves''},
\prd{42}{1990}{1960}.

\bibitem{Anto}
I.~Antoniadis,
{\it``A possible new dimension at a few TeV''},
\plb{246}{1990}{377}.

\bibitem{ADD}
N.~Arkani-Hamed, S.~Dimopoulos and G.~R.~Dvali,
{\it``The hierarchy problem and new dimensions at a millimeter''},
\plb{429}{1998}{263} [arXiv:\hepph{9803315}].

\bibitem{AADD}
I.~Antoniadis, N.~Arkani-Hamed, S.~Dimopoulos and G.~R.~Dvali,
{\it``New dimensions at a millimeter to a fermi and superstrings at a TeV''},
\plb{436}{1998}{257} [arXiv:\hepph{9804398}].

\bibitem{AbS}M.~Abramovitz and I.A.~Stegun, {\it Handbook of Mathematical
Functions}, (Dover, New York, 1965).

\bibitem{benalex}
B.~M.~N.~Carter and A.~B.~Nielsen,
{\it``Series solutions for a static scalar potential in a
Salam-Sezgin supergravitational hybrid braneworld''},
\grg{37}{2005}{1629} [arXiv:\grqc{0512024}].

\bibitem{CG}
A.~Chamblin and G.~W.~Gibbons,
{\it``Nonlinear supergravity on a brane without compactification''},
\prl{84}{2000}{1090} [arXiv:\hepth{9909130}].

\bibitem{GGP}
G.~W.~Gibbons, R.~G\"uven and C.~N.~Pope,
{\it``3-branes and uniqueness of the Salam-Sezgin vacuum''},
\plb{595}{2004}{498} [arXiv:\hepth{0307238}].

\bibitem{GM}
G.~W.~Gibbons and K.~Maeda,
{\it``Black holes and membranes in higher dimensional theories with
dilaton fields''},
\npb{298}{1988}{741}

\bibitem{MW}
S.~Mignemi and D.~L.~Wiltshire,
{\it``Spherically symmetric solutions in dimensionally reduced
space-times''},
\cqg{6}{1989}{987}

\end{thebibliography}
\end{document}